\documentclass[11pt,a4]{article}
\usepackage{times,fancyhdr}
\usepackage[dvips]{graphicx}
\usepackage{amsmath}
\usepackage{amssymb}
\usepackage{array}
\usepackage[small,bf]{caption}
\usepackage{color}
\usepackage[yyyymmdd,hhmmss]{datetime}
\usepackage{fancyhdr}
\usepackage{fancyvrb}
\usepackage{hyperref}
\usepackage{longtable}
\usepackage{natbib}
\usepackage{setspace}
\usepackage{framed}
\usepackage{titlesec}
\usepackage{titling}
\usepackage{xcolor}
\usepackage{authblk}

\fancyheadoffset[R]{0pt}
\fancyfootoffset[R]{0pt}
\definecolor{orange}{rgb}{1.0,0.5,0.0}
\definecolor{aqgr}  {rgb}{0.0,1.0,0.6} % aqua green
\definecolor{viol}  {rgb}{0.8,0.6,0.8}
\definecolor{figdr} {rgb}{1.0,1.0,1.0} % 0.8,0.6,0.7
\definecolor{colne} {rgb}{0.8,0.0,0.0} % fuchsia
\definecolor{coldr} {rgb}{1.0,0.8,0.0} % gold
\definecolor{colop} {rgb}{0.0,1.0,1.0} % blue
\definecolor{colok} {rgb}{1.0,1.0,1.0} % white

\newcommand\parok[1]{\colorbox{colok}{#1}}
\newcommand\neudef[1]{\textcolor{black}{\textit{#1}}}
\title{\vspace{0.0cm}\bfseries{\textsc{ %\huge
   Why evolution needs the old: \\
   a theory of ageing as adaptive force
}}}
\author[1]{Alessandro Fontana\thanks{fontalex00@gmail.com}}
\author[1]{Marios Kyriazis\thanks{drmarios@live.it}}
\affil [1]{National Gerontology Centre, Cyprus}

\vspace{-1.0cm} \date{}
\begin{document}

\pretitle{%
\begin{center}\LARGE
\vskip -2.2cm
\rule{\textwidth}{2.0pt}
\par
\vskip 0.5cm
}
\posttitle{
\par
\rule{\textwidth}{2.0pt}
\end{center}
\vskip -0.0cm
}

\maketitle
   
\vspace*{-0.5cm}
\begin{abstract}
At any moment in time, evolution is faced with a formidable challenge: refining the already highly optimised design of biological species, a feat accomplished through all preceding generations. In such a scenario, the impact of random changes (the method employed by evolution) is much more likely to be harmful than advantageous, potentially lowering the reproductive fitness of the affected individuals. Our hypothesis is that ageing is, at least in part, caused by the cumulative effect of all the experiments carried out by evolution to improve a species' design. These experiments are almost always unsuccessful, as expected given their pseudorandom nature, cause harm to the body and ultimately lead to death. On the other hand, a small minority of experiments have positive outcome, offering valuable insight into the direction evolution should pursue. This hypothesis is consistent with the concept of ``terminal addition'', by which nature is biased towards adding innovations at the end of development. From the perspective of evolution as an optimisation algorithm, ageing is advantageous as it allows to test innovations during a phase when their impact on fitness is present but less pronounced. Our inference suggests that ageing has a key biological role, as it contributes to the system's evolvability by exerting a regularisation effect on the fitness landscape of evolution.
\end{abstract}

\section{Introduction}

% introduction
\parok{Ageing is a process}that occurs in the lifespan of most living beings. It involves the accumulation of changes in the organism over time, leading to a progressive deterioration of bodily functions. Ageing does not affect all species: in some species, its effects are negligible or undetectable. Moreover, the rate of the process is very different among affected species: a mouse is old at three years of age, a human at eighty. Some species (e.g., sturgeon, turtles, some mollusks, lobsters) show no age-related mortality increase in the wild \citep{Finch90, Finch01}.

% ageing facts
\parok{Numerous genes have been}demonstrated to influence lifespan in certain species. Sets of genes that are co-expressed during youth gradually lose their correlation in expression over time \citep{Southworth09}: in other words, the level of randomness in gene regulatory networks seems to increase with age progression. An effective measure that has been shown to reduce the rate of ageing in model organisms and rodents is calorie restriction without malnutrition \citep{Al-Regaiey16}, with beneficial effects on longevity and on the incidence of age-related diseases such as diabetes, cardiovascular diseases and cancer, for which old age represents the biggest risk factor \citep{deMagalhaes13}.

% stochastic and programmed theories
\parok{Numerous ageing theories have}been proposed. \textit{Stochastic theories} attribute ageing to damage incurred by organisms due to environmental factors or products of cellular metabolism (free radicals), akin to the ageing of mechanical devices. According to \textit{programmed theories}, ageing is steered by a set of instructions encoded in DNA which can, to some extent, be influenced by environmental conditions. \textit{Epigenetic ageing theories} \citep{Yang23epinfo} propose that changes in gene expression patterns, rather than modifications to the DNA sequence itself, underlie age-related shifts in cellular function and tissue degeneration.

% ageing and evolution
\parok{Despite extensive investigation, the}biological significance of ageing remain enigmatic. Even more perplexing is the intricate relationship between ageing and evolution, the force that shapes biological systems. The influence of evolution on the ageing profile of a species is rooted in the established notion that the evolutionary pressure on an individual diminishes as age progresses. The \textit{disposable soma theory} \citep{Kirkwood77, Kyriazis17} states that organisms age due to an evolutionary trade-off between growth, reproduction, and DNA repair maintenance, possibly mediated by the phenomenon of \textit{antagonistic pleiotropy} \citep{Williams57}, by which a single gene influences multiple traits, with benefits early in life but detrimental effects later on.

% ageing and evolution
\parok{The concept of}\textit{evolution of ageing}, reviewed in \citep{Fabian11}, centers on the notion that ageing is an adaptive phenomenon that has evolved because it provides certain advantages either to individual organisms or to the species as a whole. The notion of ageing as a purposeful, programmed sequence of events holds an inherent appeal due to its numerous conserved aspects and the demonstrated potential to alter lifespan by manipulating individual genes or pathways. However, the proposition of ageing as a nonadaptive phenomenon is robust and nearly universally acknowledged within the field \citep{Vijg16essence}.

% ET model intro
\parok{This work is based}on the assumption that ageing is not merely a side-effect of genetic traits advantageous during development, but is (mostly) actively driven by a program under genetic control, that starts at the zygote stage and proceeds throughout the entire lifespan. This is the main idea behind a model called \neudef{Epigenetic Tracking (ET)} \citep{Fontana08}, capable of devo-evolving complex 3d structures in computer simulations. ET was originally conceived to model development, and later extended to interpret a range of biological phenomena \citep{Fontana12b, Fontana23a}, including ageing \citep{Fontana14}. 

% our hypothesis
\parok{As described in previous}work, ET is consistent with the notion that evolution affects ageing through forces that decrease with age progression. In this work we turn our attention on the reciprocal effect of ageing on evolution. By framing evolution as an optimisation algorithm and approaching the issue through the lens of the principle of terminal addition, we will present a hypothesis proposing that ageing exerts a beneficial influence on evolution, playing a key biological role. It is worth emphasising that our proposal acknowledges the complexity of the ageing phenomenon, and does not discount the potential involvement of other factors, such as environmental damage, in this process.

\section{The evo-devo model}

\begin{figure*}[t] \begin{center} \hspace*{-0.50cm}
\includegraphics[width=18.00cm]{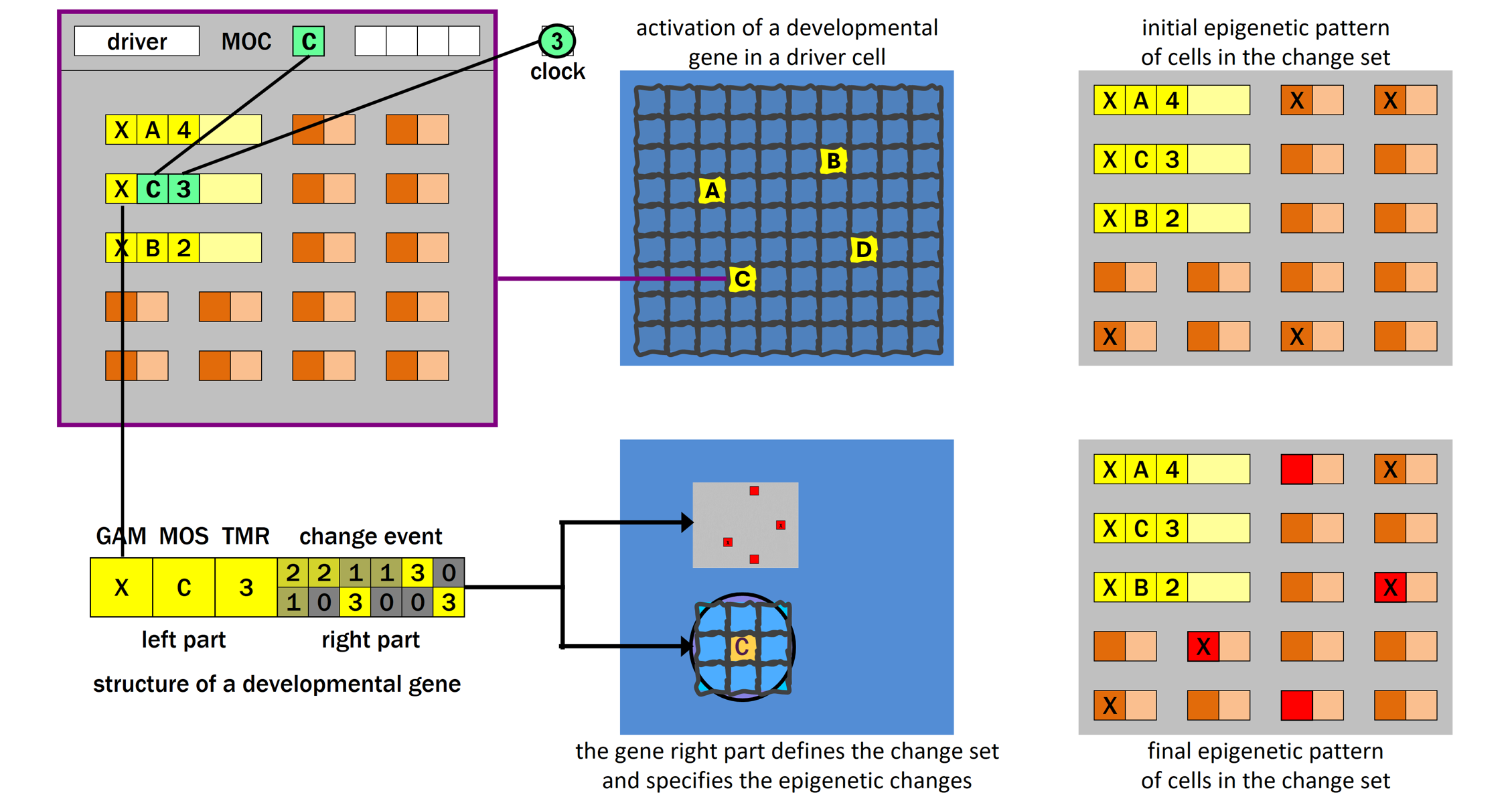}
\caption{
\parok{A change event of}type differentiation. In driver cell C (upper left and upper middle panels) a developmental gene is activated. The code in the gene right part (lower left panel) specifies the change set and the epigenetic changes (lower middle panel) to be applied to the initial epigenetic pattern (upper right panel). In the final epigenetic pattern (lower right panel), the availability marks of some genes are modified: some genes may be made available, some may get disabled.}
\label{events}
\end{center} \end{figure*}

% phenotypes, cells
\parok{In this model, phenotypes}are represented by cellular formations composed of cube-shaped cells organised on a grid. These cells can be categorised into two groups: \neudef{normal cells}, which constitute the majority of the structure, and \neudef{driver cells}, which are considerably fewer in quantity and uniformly dispersed throughout the structure's volume. Normal cells could be likened to basic soldiers, while driver cells might be envisioned as captains due to their ability to command normal cells. The outcome of these commands is the orchestration of \neudef{change events}, encompassing the generation, elimination, and specification of significant numbers of cells. Events are synchronised through a global clock shared by all cells.

% genome and epigenome
\parok{Cells have a genome,}with two categories of genes: \neudef{developmental genes}, which regulate development; \neudef{other genes}, which carry out all other functions. Besides the genome, which is the same for all cells, there are also elements that can vary between cells, reflecting the differentiation process that occurs during development. These include the \neudef{driver mark}, which distinguishes between driver and normal cells, and the \neudef{master organisation code (MOC)}, an abstraction for all master regulatory elements in the cell. The biological analogous of the variable elements may include elements from the epigenome, as well as from the transcriptome. 

% develop genes /left part
\parok{Developmental genes can be}likened to ``macro'' sets of genes that are regulated together. The left part of a developmental gene (lower left panel of Fig.~\ref{events}) is composed of three elements: i) the \neudef{gene availability mark (GAM)} that determines if the gene is available for activation or structurally inactive; ii) the \neudef{master organisation sequence (MOS)} that can align with the MOC; iii) the \neudef{timer (TMR)} that can match with the clock. MOC and MOS sequences are symbolised by letters, TMR and clock are shown as numbers. At each developmental stage, identified by a distinct clock value, the MOC and MOS sequences are compared for each developmental gene and driver cell. When the MOS aligns with the MOC and the timer matches the clock, the gene becomes active and the gene's right part is executed.

% biological interpretation
\parok{The MOC and the}clock are responsible for triggering the activation of various parts of the genetic code for development in a specific spatial and temporal manner. In biological systems, the MOC may correspond to the cell master regulatory elements, able to initiate a chain reaction of gene activations that determines the cell behaviour. The clock could be implemented as a diffusible molecule, emitted by a central source and propagated to all cells. The GAM's may correspond to epigenetic information (e.g. histone modifications or methylation marks) that renders genes either available for activation or structurally inactive.

% develop genes /right part
\parok{The right part of}a developmental gene encodes a \neudef{change event}. Three types of events are foreseen. \neudef{Differentiation change events} (Fig.~\ref{events}) cause the activated driver cell to switch the availability marks of some genes in a set of cells called \neudef{change set}), which can be so small as to include only the driver itself. These events may correspond to writing, erasing and reading operations within the ``histone code'' framework \citep{Jenuwein01}. \neudef{Proliferation change events} cause the activated driver cell to proliferate (or induce the proliferation of other cells). The outcome of \neudef{apoptosis change events}, which mimics the phenomenon of biological apoptosis, is the removal of cells in the change set from the grid. 

% generation of new driver cells
\parok{Most normal cells generated}during a proliferation event become terminally differentiated cells. This process may be implemented in nature through a class of molecules called growth factors, which during embryonic development act locally as important regulators of cellular proliferation and differentiation \citep{Ruddon09}. Some other cells, instead of (or in addition to) proceeding on their differentiation path, revert to driver cells. Hence, in ET the normal-driver transition is a two-way street and may correspond to a fundamental principle present also in nature \citep{Gupta11}, that induces cellular systems to self-organise in a hierarchical fashion. In ET the process of normal-driver conversion is achieved through the presence of simulated morphogens \citep{Fontana13a}. 

% generation of new MOC
\parok{The newly formed driver}cells derived from normal cells acquire a new and distinct MOC value to permit differentiation. Should a developmental gene exist in the genome with a MOS corresponding to the MOC of a new driver cell, this cell can serve as the focal point for another event in a subsequent stage, propelling the developmental progression. Thus, development foresees the alternation of two phases: a phase A, where driver cells produce new normal cells and induce their differentiation, and a phase B, where new driver cells emerge from normal cells. Continuously repeated, this cycle forms the core of development. %In Figure~\ref{organdev}, an illustration of development for a hypothetical ``organ'' demonstrates the alternating sequence of change events and the generation of new driver cells.

% Map stem /driver
\parok{The significance of driver}cells for development in ET raises a natural question about the relationship between driver cells and stem cells. The term ``stem cell'' can have two distinct meanings in biology. Firstly, it can refer to cells that have the ability to \emph{induce} other cells to undertake specific actions. Secondly, it can refer to cells that can \emph{be induced} to differentiate into specific cell types due to their plasticity. In line with this distinction, we have chosen to reserve the term ``stem cell'' for the second meaning, and instead use the term ``driver cells'' to refer to the first meaning.

%\section{The evo-devo process}
 
\begin{figure}[t] \begin{center} \hspace*{-0.20cm}
\includegraphics[width=17.40cm]{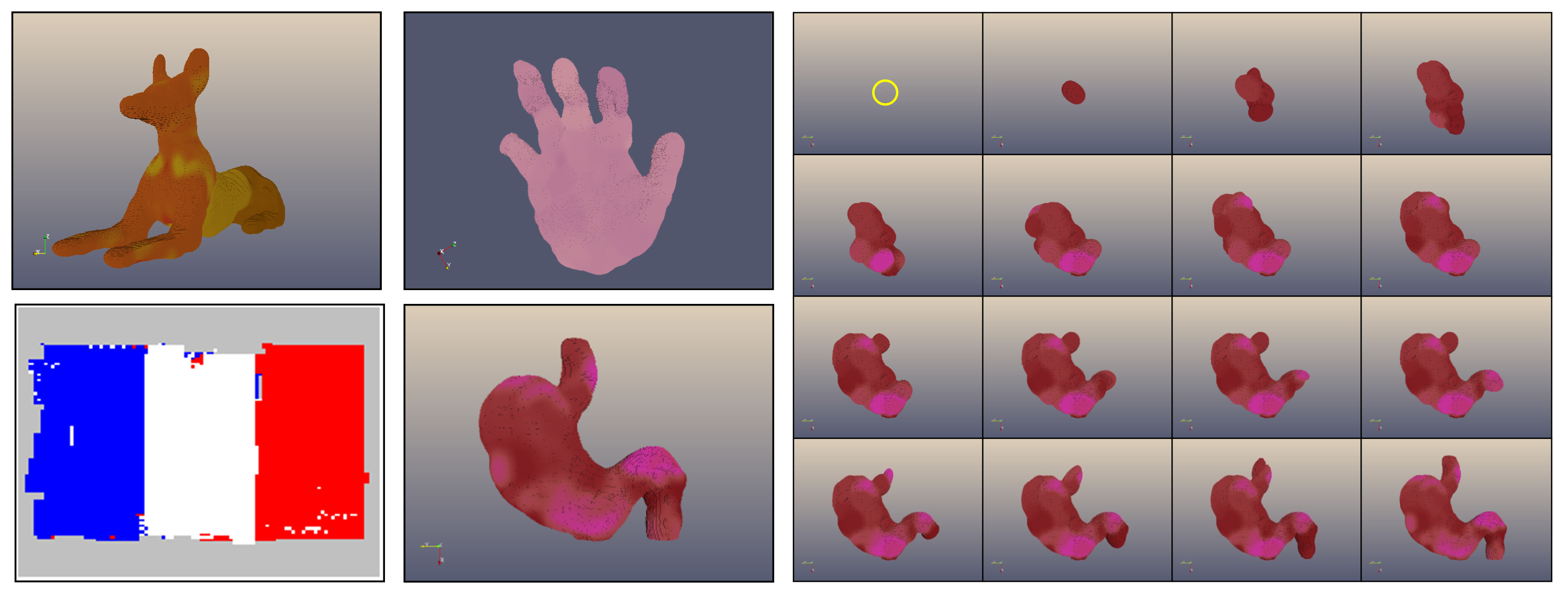}
\caption{
On the left: \parok{some devo-evolved 3-dimensional}coloured structures. On the right: development sequence of a structure representing a human stomach.}
\label{stomach}
\end{center} \end{figure}

% genetic algorithm
\parok{The development model described}can be integrated with a \textit{genetic algorithm} that emulates biological evolution. In this algorithm, a population of individuals, each encoded in an artificial genome, undergoes evolution over multiple generations. During each generation, all individuals independently progress from the zygote stage to the final phenotype. The proximity of the final phenotype to a predefined target is used as a fitness measure. This process is repeated for all individuals, resulting in the assignment of a fitness value to each. Based on these values, individuals' genomes are selected and subject to random mutations, generating a new population. This iterative cycle continues until a satisfactory level of fitness is achieved.
 
% evo-devo method and results
\parok{The coupling of the}model of development and the genetic algorithm gives origin to an evo-devo process aimed at generating 3-dimensional structures. In computer simulations, or ``in silico'' experiments (refer to examples in Fig.~\ref{stomach}), this process has demonstrated its effectiveness in ``devo-evolving'' structures of unparalleled complexity, encompassing both shape and function \citep{Fontana10b}. The efficacy of this approach hinges largely on the characteristics of the development model, particularly the uniform distribution of driver cells, which keeps the structure plastic throughout development, enabling the construction of new substructures on top of existing ones.

%\clearpage
%~\newpage

\section{Ageing as a program driven by timer-dependent events}

% clock beyond reproduction
\parok{This section provides a}summary of the interpretation of ageing according to ET exposed in \citep{Fontana14, Fontana23a}. In this model, an individual undergoes development across N stages, culminating in the evaluation of their fitness, which is then employed for selection in the genetic algorithm. In most experiments, the moment of fitness evaluation coincided with the end of the simulation. However, an alternative scenario involves allowing the global clock to continue ticking, observing the unfolding events post-fitness evaluation. This distinction between the periods before and after fitness evaluation mirrors the biological phases of development (e.g., up to 25 years of age in humans) and ageing (from 25 years of age onward). 

% nature of ageing
\parok{At the end of}development, many driver cells are present in the body of the individual. Some of these driver cells have been activated during development (and shaped the individual's body), some (the vast majority in previous simulations) have not. These inactive driver cells may contain developmental genes destined to activate during the ageing period, occurring after the fitness evaluation moment when, by definition, they do not influence the fitness value. Consequently, their right parts have not been optimised through evolution, leading to associated events exhibiting a ``pseudorandom'' nature, i.e. they appear random despite being encoded in the genome, subject to genetic control, and deterministic in essence. The implication of this pseudorandomness is that these events are more likely to have detrimental effects on the individual than beneficial ones. 

\begin{figure}[t] \begin{center} \hspace*{-0.00cm}
\includegraphics[width=15.00cm]{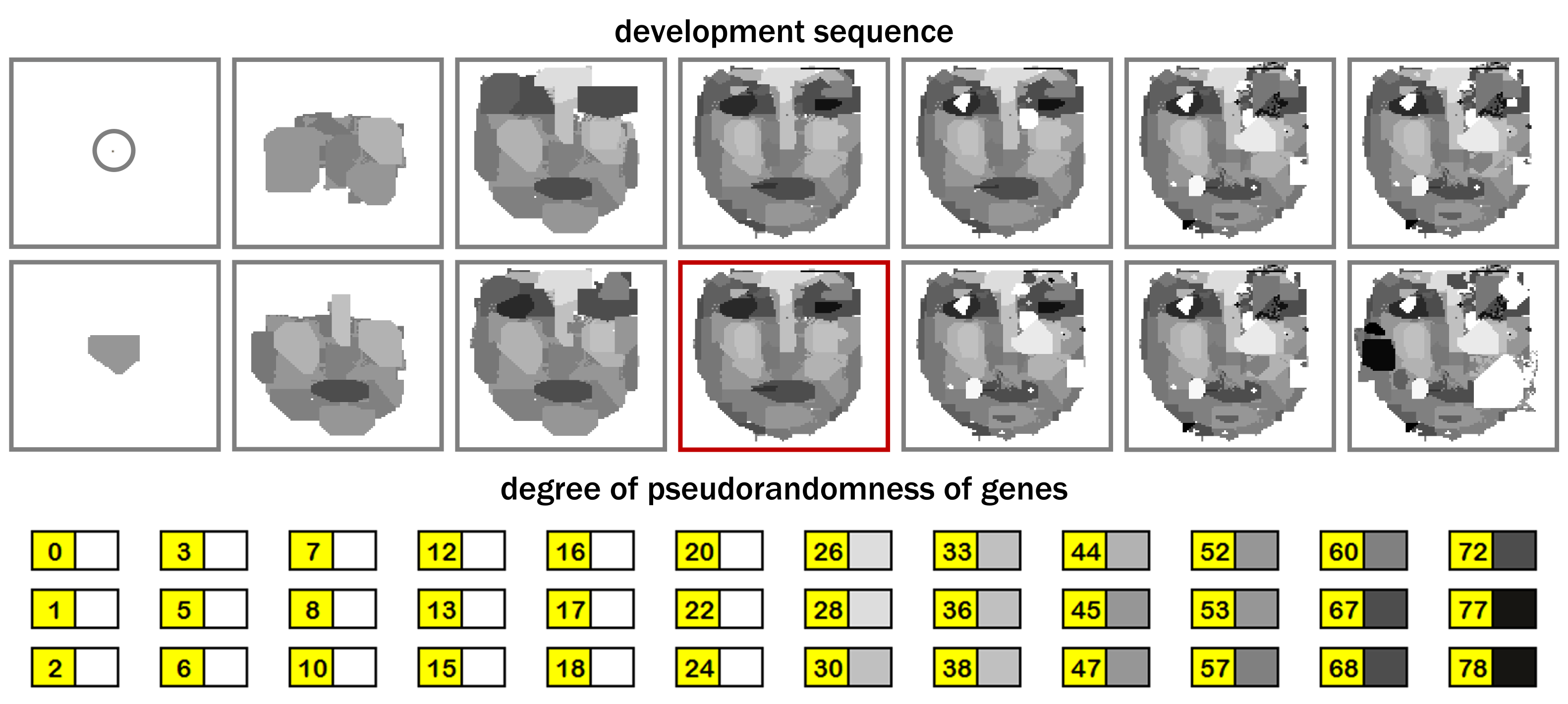}
\caption{
\textbf{Upper panel.} \parok{Simulation of ageing for}a ``face''. On the left the period of development: the structure grows from a single cell to the mature phenotype at age 25 (red frame), when fitness is evaluated. On the right the period of ageing: the quality of the structure deteriorates under the action of non-optimised events. \textbf{Lower panel.} The evolutionary pressure acting upon a developmental gene (captured by the gradient $\lvert \nabla F(w_i) \rvert$) is an inversely proportional function of the gene timer value. The opposite is true for the level of pseudorandomness, represented by the grey shade. Genes with timer values $\leq 25$ are subject to a high pressure, genes with timer values $\geq 25$ are subject to a steadily decreasing pressure, until their effects become completely pseudorandom.}
\label{ageface}
\end{center} \end{figure}

% gradual reduction of pressure
\parok{The evolutionary pressure acting}on a developmental gene can be captured by the absolute value of the gradient of fitness F with respect to the parameters the gene depends upon: $\lvert \nabla F(w_i) \rvert$. The hypothesis that posits the complete absence of pressure on a gene activated post-reproduction is oversimplified. In nature, an individual's reproductive fitness is influenced by events occurring after reproduction (e.g. the quality of care of parents and grandparents), as also these events can impact the survival chances of the offspring. A more realistic perspective suggests that the effect of an event on fitness, and consequently the pressure on the corresponding gene, tends to diminish as the age of its manifestation (determined by the associated timer value) increases, rather than abruptly vanishing immediately after reproduction. 

% painter analogy
\parok{As highlighted in section}1, the concept that the evolutionary pressure wanes with age progression is not novel. A distinctive feature of the ET model is interpreting this phenomenon as directed by an \textit{active program}, rather than simply arising as a byproduct of effects beneficial during early life. To illustrate, imagine the developmental process like a robotic painter following genome-encoded instructions to apply brushstrokes. Each gene defines the characteristics of a stroke, such as timing, shape, colour, and texture. The process starts with an empty canvas and advances gradually to craft the complete artwork. 

% painter analogy
\parok{Drawing on this analogy,}a common implicit assumption in most ageing theories is that the painter continues working until the end of development, at which point other mechanisms take over (like colours diffusing or degrading), causing the painting to become blurred and deteriorated. However, the ET model posits that the painter never stops: it remains active from the moment of conception until the end of life. However, the brushstrokes, highly optimised and fine-tuned during development, progressively become more random-like, as a result of the declining pressure. Figure~\ref{ageface} shows a simulation of ageing for a ``face'', incorporating this insight.
 
% ageing driver cells
\parok{Therefore, ageing is interpreted}as a \emph{continuation of development}, driven by non-optimised genes activated in specific driver cells after reproduction. There is empirical evidence of an age-related decline in the functionality of adult stem cells \citep{Liu11}. Our hypothesis is that this functional decline is determined by epigenetic changes orchestrated in biological driver cells by change events associated to timer values, bound to occur at precise moments. Such events may occur in driver cells activated during development, and reactivated in the post-reproductive period for tissue repair, or in driver cells activated in the post-reproductive period for the first time. These predictions are consistent with the reduced epigenetic coherence observed in cells, that constitutes one of the hallmarks of the ageing phenotype \citep{deMagalhaes23b, Yang23epinfo}.
 
% ageing-related diseases
\parok{Viewing ageing as the}gradual accumulation of pseudorandom events offers an elegant explanation for diseases typically associated with old age, such as Alzheimer's disease or type II diabetes. In this perspective, the difference between the phenotipic manifestations of such diseases and the effects of ``normal'' or ``healthy'' ageing are more quantitative than qualitative. Both normal ageing and ageing-associated diseases are driven by the same underlying mechanism: the manifestations of normal ageing are simply milder and more benign. This notion finds strong support in the fact that measures known to delay ageing, such as calorie restriction, also postpone the onset of age-related diseases \citep{Colman09}. 
 
\section{Ageing as evolution in action}

%\begin{figure}[t] \begin{center} \hspace*{-0.25cm}
%\includegraphics[width=17.50cm]{e01evolsets}
%\caption{
%\if\afhead1 {\parop{figurex}} \fi
%\if\afhead1 {\parop{kaption}} \fi
%\textbf{Genetic exchange without ageing.} Developmental genes can be divided in genes activated during life (L), and genes never activated during life (J). Elements are free to move between sets: the set of inactive elements represent therefore a reservoir of potential new events. \textbf{Genetic exchange with ageing.} Ageing provides a buffer zone where new solutions can be tried with limited impact on evolutionary fitness.}
%\label{evolsets}
%\end{center} \end{figure}

% validity beyond ET
\parok{The theory of ageing}presented hereafter is situated within the context of the ET evo-devo model. Nevertheless, its relevance extends beyond this framework to encompass any evo-devo model where developmental processes are governed by programmed, time-dependent events. A fundamental prerequisite is that evolution possesses the ability to modify the timing of these events, either delaying or advancing their onset. Specifically, this hypothesis is consistent with programmed theories of ageing, positing that ageing stems from events with associated timers under genetic control, which evolution can adjust.

% evolution task
\parok{As an optimisation algorithm,}evolution continuously confronts a formidable task: refining the design of biological species. This iterative process has persisted for hundreds of millennia, resulting for most species in a highly optimised biological blueprint. In such a scenario, the impact of random changes (the method employed by evolution) is much more likely to be harmful than advantageous for the affected individuals, potentially lowering their reproductive fitness. Hence, the effects of various detrimental changes should manifest in all individuals of evolving species. 

% causes and effects
\parok{One crucial question arises:}where do these effects manifest? The streamlined code that drives development serves as evidence of past evolution. Yet, what about present evolution? It appears to be a \textit{cause without an effect}. The existence of these effects can be excluded by invoking the intervention of external forces (e.g. intelligent design), that would dramatically reduce the number of unsuccessful experiments. On the other hand, there is a phenomenon ---ageing---  that appears to be driven by the accumulation of random-like, deleterious events, although the molecular details are still unclear, as well as the interplay between environmental and genetic factors. In this case we have an \textit{effect without a cause}. 

% ageing is the answer
\parok{Hence, there is an}opportunity to catch two birds with one stone, suggesting that evolution is the cause of ageing and ageing is the effect of evolution. We argue that ageing is, at least to some extent, caused by the cumulative effect of all the experiments carried out by evolution to improve a species' design. These experiments are almost always unsuccessful, as expected given their random-like nature, cause harm to the body and ultimately lead to death. On the other hand, a small minority of experiments have positive outcome, offering valuable insight into the direction evolution should pursue. This does not rule out the possibility that other factors, such as environmental damage, may also play a role in the ageing process.

% contrast with disposable soma
\parok{This proposal sharply contrasts}with the disposable soma theory, which posits that ageing is the result of an increasingly absent evolution. Our hypothesis is that ageing, essentially, \textit{is} evolution. This raises another question: why aren't innovations integrated directly into the developmental phase, where their impact on fitness would be maximal? We speculate that nature prefers conducting experiments after reproduction, during the later stages of an individual's life. An elucidation of this phenomenon is also required.

% evol sets
\parok{In ET, developmental genes}can be divided in i) genes activated during development, ii) genes activated during ageing and iii) genes never activated. On the other hand, elements are free to move between sets: the sets of inactive elements represent therefore a reservoir of potential new events and a tool to explore new evolutionary paths. One potential scenario involves the flow of change events from set iii) to set i), maximising fitness impact. However, as observed, this approach carries a high risk of disrupting an already fine-tuned developmental trajectory. Set ii) (ageing) can provide a buffer zone, a phase where the impact on fitness is still present but diminished. This configuration offers a safer environment for the introduction of new change events, enabling experimentation with novel solutions with a reduced risk.

% terminal addition
\parok{This hypothesis is consistent}with the concept of ``terminal addition'' \citep{Gould77} or ``peramorphosis'' \citep{Alberch79}. This principle, which represents the modern version of the ``ontogeny recapitulates phylogeny'' theory originally proposed by Ernst Haeckel, is commonly conceptualised  as the addition of a new developmental stage (an innovation) at the end of development. The hypothesis is grounded in the idea that alterations to a developmental trajectory at a specific stage are constrained by the structure established in preceding stages. Modifications impacting earlier stages are anticipated to have a more profound influence on the final structure compared to changes occurring later. Consequently, it is reasonable to expect that nature is biased towards adding innovations at later stages.

% decrease of timer values
\parok{In a hypothetical scenario,}a new change event could emerge, with effects manifesting at old age. The impact on fitness will be small but present: if positive, the innovation will be accepted and incorporated in the design of future generations, slowly spreading in the population. After confirming the ``safety'' of the change during the ageing period, evolution could manipulate the corresponding timer value, shifting it towards lower values within the earlier ageing period where its influence on fitness is greater. Upon validating the safety of the innovation at an earlier phase, it becomes possible to shift its occurrence even further, allowing it to reach the developmental period. The final outcome is a more efficient evolutionary process, capable of selecting favorable traits without excessively disrupting the existing design.

% analogy with drug clinical trials
\parok{An analogy can be}drawn with drug clinical trials, which progress through several phases, each with a specific focus. Phase I trials involve a small number of healthy volunteers to assess the drug's safety profile, dosage range, and potential side effects. If successful, the drug advances to Phase II, where a larger group of patients with the target condition participates to further evaluate its effectiveness and side effects. In Phase III trials, the drug is tested on an even larger scale, often involving thousands of participants, to confirm its benefits, monitor side effects, and compare it to existing treatments. This scheme is used to test new interventions within a confined setting, where the potential harm is minimised. We argue that the ageing period serves an analogous purpose for biological innovations brought about by natural selection.

% disposable vs evolvable soma
\parok{Based on this hypothesis,}the pseudorandomness of gene expression depicted in Figure~\ref{ageface} is not just an incidental byproduct of diminishing evolutionary pressure, but becomes an opportunity to explore the space of innovations. Once an individual has fulfilled its reproductive duties, its body is not disposed of (and wasted), but used for experimentation by evolution. This is done initially with caution, as evolution is aware that the individual still has parental tasks to fulfill, and increasingly more aggressively as the offspring becomes less dependent on parental care. According to this hypothesis, ageing can be viewed as nothing less than \textit{evolution in action}. 

\section{Ageing shapes the fitness function of evolution}

\begin{figure}[t] \begin{center} \hspace*{-0.20cm}
\includegraphics[width=17.40cm]{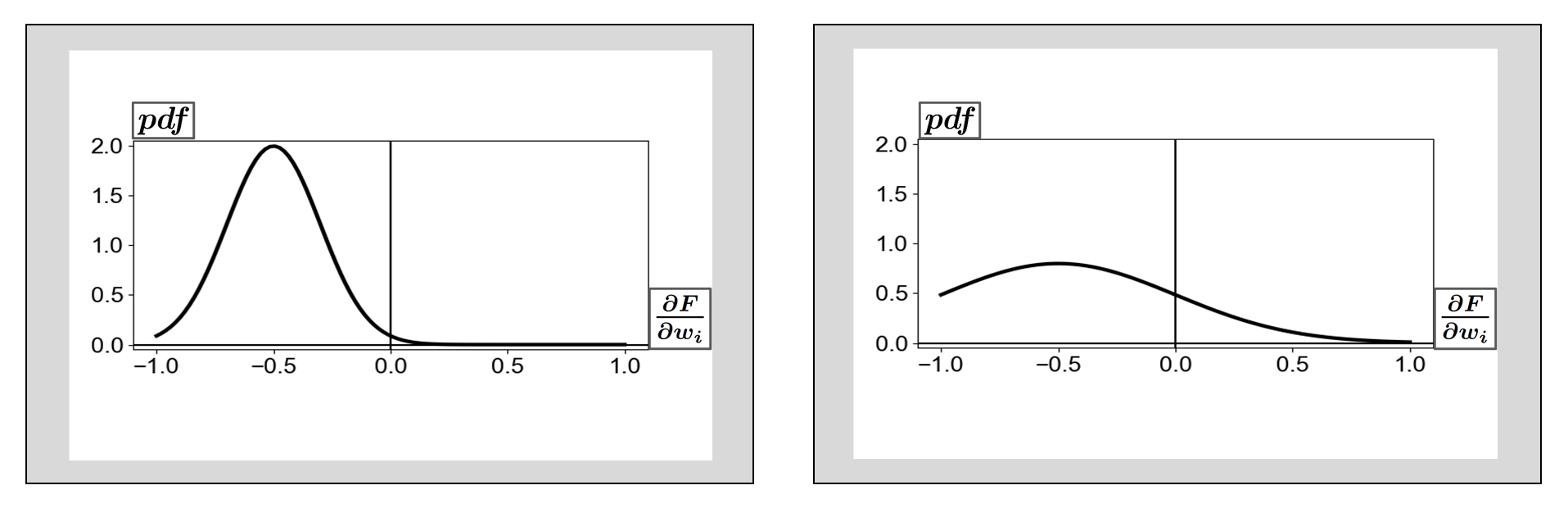}
\caption{
\parok{Probability density functions (pdf)}of fitness impact of modifications to the parameters that encode developmental genes, formalised as the partial derivative $\frac{\partial F} {\partial w_i}$ of the fitness function F with respect to a generic parameter $w_i$. The extension of development with ageing spreads the impact on fitness of a parameter change to a wide range of values. This translates the initial distribution (left) into a new distribution (right) with higher variance and a heavier right tail.}
\label{pdfcurves}
\end{center} \end{figure}

% introduction of pdf
\parok{The proposed phenomenon can}be characterised in a more quantitative manner by considering the probability distribution of the evolutionary impact resulting from modifications to the parameters that encode developmental genes. The probability distribution is captured by the probability density function ($pdf$) and the said impact can be formalised as the partial derivative $\frac{\partial F} {\partial w_i}$ of the evolutionary fitness function F with respect to a generic parameter $w_i$. We assume that the initial $pdf$ exhibits a Normal shape centered around -0.5 (see Figure~\ref{pdfcurves}, left). This choice is consistent with our earlier assumption that the effect of a random mutation is more likely to be harmful than advantageous.

% smoother landscape
\parok{The extension of development}(having a high and constant evolutionary pressure, e.g. equal to 1) with ageing (characterised by a decreasing pressure, comprised e.g. between 0 and 1) has the effect of spreading the aforementioned impact on fitness to a broader range of values. This translates the initial effect distribution into a new distribution with higher variance (Figure~\ref{pdfcurves}, right) and a heavier right tail, thereby increasing the chances of a positive outcome. The increased probability density of positive impacts offers evolution more possibilities to explore the space of parameters and find good design innovations. Therefore, it can be expected that most successful innovations are introduced during later life stages, throughout the aging period.

\begin{figure}[t] \begin{center} \hspace*{-0.00cm}
\includegraphics[width=17.00cm]{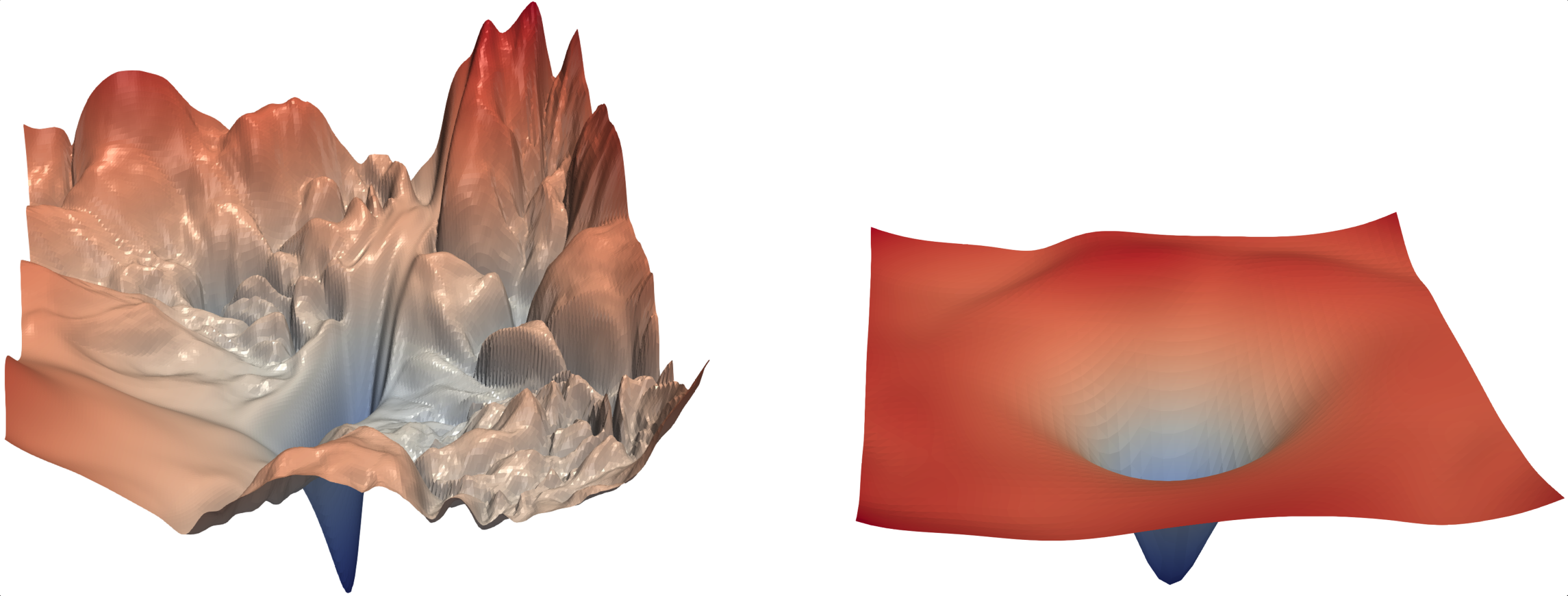}
\caption{
\parok{The shape of the}loss function of a deep neural network without (left) and with (right) skip connections \citep{Li17landscape}. The introduction of skip connections promotes the regularisation of the shape of the loss function, optimised by the gradient descent algorithm. The ageing period is hypothesised to have a similar smoothing effect on the shape of the evolutionary fitness function, optimised by the force of natural selection, simulated with a genetic algorithm.}
\label{landscape}
\end{center} \end{figure}

% fitness landscape for SGD
\parok{A smooth fitness landscape}plays a crucial role in optimisation by gradient descent, the algorithm employed to train modern AI systems. The structure of the relationship between a model's parameters and its performance metric determines the efficiency and effectiveness of gradient-based optimisation algorithms. In a smooth landscape, incremental changes in parameter adjustments lead to predictable shifts in performance, enabling the algorithm to navigate confidently towards optimal solutions. With reference to neural networks (the most commonly used class of AI models), one of the benefits of skip connections consists in regularising the shape of the loss function (Figure~\ref{landscape}, \cite{Li17landscape})

% fitness landscape for GA
\parok{The landscape's smoothness or}ruggedness impacts the quality of the optimisation process also in the context of genetic algorithms, which represent the algorithmic version of biological evolution \citep{Malan21}. In this scenario as well, a smooth fitness landscape generally implies that small changes in an individual's genetic makeup lead to relatively predictable changes in its fitness (or objective) value. In such cases, the genetic algorithm may converge more quickly and effectively towards optimal solutions, as the evolutionary steps can be more controlled and directed. This captures, according to our hypothesis, the effect of ageing on evolution.

\section{Conclusions and future work}

% recap
\parok{In this work we}have proposed a novel biological hypothesis, suggesting that evolution is the cause of ageing and ageing is the effect of evolution. We contend that ageing is, at least partially, a consequence of the cumulative impact of evolutionary experiments aimed at enhancing a species' design. These experiments, typically unsuccessful due to their random nature, inflict harm on the body and ultimately inevitably lead to death. Our proposition acknowledges the potential involvement of other factors, such as environmental damage, in the ageing process.

% recap
\parok{Moreover, we suggest that} ageing enhances the system's evolvability by exerting a regularisation effect on the evolutionary fitness landscape. The incorporation of an ageing period provides evolution with a buffer zone, where design innovations can be tested, with a lower impact on fitness and a reduced risk. In this context, evolution would still be able to see the optimal direction of parameter change (the gradient), an information that can be exploited to further refine the innovations in subsequent generations. This translates to an increased effectiveness of the evolutionary process.

% lifespans of living fossils
\parok{Future endeavors will focus}on substantiating the proposed hypothesis. If ageing indeed proves advantageous for evolution, we might conjecture a direct correlation between a species' evolvability and its susceptibility to ageing. This relationship could be investigated by examining lifespans and evolutionary stability across various species. Given the vast spectrum of lifespans observed in biology, meticulous attention to detail is required to ensure accurate comparisons between similar species and the exclusion of spurious effects from the analysis.

% simulations
\parok{A second approach we}are planning to undertake will seek to explore the relationship between evolution and ageing through computer simulations, utilising both the ET model and other platforms. This is the approach of Artificial Life \citep{langton1989}, a field situated at the intersection of biology, computer science, and engineering. In this context, research objectives may shift from merely understanding biology to constructing biologically inspired artifacts that exhibit characteristics akin to those of biological organisms, such as complexity or resilience to damage. The hope is that these characteristics will emerge if the artifacts are engineered using bioinspired methodologies, including evolution and development.

\bibliographystyle{apalike}
% or: plain,unsrt,alpha,abbrv,acm,apalike,...
\bibliography{lxevolager}

\begin{thebibliography}{}

\bibitem[Al-Regaiey, 2016]{Al-Regaiey16}
Al-Regaiey, K.~A. (2016).
\newblock The effects of calorie restriction on aging: a brief review.
\newblock {\em Eur Rev Med Pharmacol Sci}, 20(11):2468--73.

\bibitem[Alberch et~al., 1979]{Alberch79}
Alberch, P. et~al. (1979).
\newblock Size and shape in ontogeny and phylogeny.
\newblock {\em Paleobiology}, 5:296--317.

\bibitem[Colman et~al., 2009]{Colman09}
Colman, R.~J. et~al. (2009).
\newblock Caloric restriction delays disease onset and mortality in rhesus
  monkeys.
\newblock {\em Science}, 325(5937):201--204.

\bibitem[de~Magalh\~{a}es, 2013]{deMagalhaes13}
de~Magalh\~{a}es, J.~P. (2013).
\newblock How ageing processes influence cancer.
\newblock {\em Nature Reviews Cancer}, 13:357--365.

\bibitem[Duffield et~al., 2023]{deMagalhaes23b}
Duffield, T. et~al. (2023).
\newblock Epigenetic fidelity in complex biological systems and implications
  for ageing.
\newblock {\em bioRxiv}.

\bibitem[Fabian and Flatt, 2011]{Fabian11}
Fabian, D. and Flatt, T. (2011).
\newblock The evolution of aging.
\newblock {\em Nature Education Knowledge}, 3(10):9.

\bibitem[Finch, 1990]{Finch90}
Finch, C. (1990).
\newblock {\em Longevity, Senescence, and the Genome}.
\newblock University of Chicago Press.

\bibitem[Finch and Austad, 2001]{Finch01}
Finch, C. and Austad, S. (2001).
\newblock History and prospects: symposium on organisms with slow aging.
\newblock {\em Experimental Gerontology}, 36:593--597.

\bibitem[Fontana, 2008]{Fontana08}
Fontana, A. (2008).
\newblock Epigenetic tracking, a method to generate arbitrary shapes by using
  evo-devo techniques.
\newblock In {\em Proceedings of the international conference on Epigenetic
  Robotics (EPIROB)}.

\bibitem[Fontana, 2010]{Fontana10b}
Fontana, A. (2010).
\newblock Devo co-evolution of shape and metabolism for an artificial organ.
\newblock In {\em Proceedings of the international conference on the simulation
  and synthesis of living systems (ALIFE)}, pages 15--23.

\bibitem[Fontana, 2023]{Fontana23a}
Fontana, A. (2023).
\newblock Unravelling the nexus: Towards a unified model of development,
  ageing, and cancer.
\newblock {\em Biosystems}, 2023 Sep;231:104966.

\bibitem[Fontana and Wr\'obel, 2012]{Fontana12b}
Fontana, A. and Wr\'obel, B. (2012).
\newblock A model of evolution of development based on germline penetration of
  new "no-junk" {DNA}.
\newblock {\em Genes}, 3:492--504.

\bibitem[Fontana and Wr\'obel, 2013]{Fontana13a}
Fontana, A. and Wr\'obel, B. (2013).
\newblock An artificial lizard regrows its tail (and more): regeneration of
  3-dimensional structures with hundreds of thousands of artificial cells.
\newblock In {\em Proceedings of the {E}uropean conference on Artificial Life
  (ECAL)}.

\bibitem[Fontana and Wr\'obel, 2014]{Fontana14}
Fontana, A. and Wr\'obel, B. (2014).
\newblock Pseudorandomness of gene expression: a new evo-devo theory of ageing.
\newblock {\em Current aging science}, 7:48--53.

\bibitem[Gould, 1977]{Gould77}
Gould, S.~J. (1977).
\newblock {\em Ontogeny and phylogeny}.
\newblock {Harvard University Press}.

\bibitem[Gupta et~al., 2011]{Gupta11}
Gupta, P.~B. et~al. (2011).
\newblock Stochastic state transitions give rise to phenotypic equilibrium in
  populations of cancer cells.
\newblock {\em Cell}, 146(4):633--644.

\bibitem[Jenuwein and Allis, 2001]{Jenuwein01}
Jenuwein, T. and Allis, C.~D. (2001).
\newblock Translating the histone code.
\newblock {\em Science}, 293:1074--1080.

\bibitem[Kirkwood, 1977]{Kirkwood77}
Kirkwood, T.~B. (1977).
\newblock Evolution of ageing.
\newblock {\em Evolution}, 270(5635):301--304.

\bibitem[Kyriazis, 2017]{Kyriazis17}
Kyriazis, M. (2017).
\newblock The indispensable soma hypothesis in aging.
\newblock In Ahmad, S.~I., editor, {\em Aging, exploring a complex phenomenon},
  chapter~4. CRC Press.

\bibitem[Langton, 1989]{langton1989}
Langton, C. (1989).
\newblock {\em Artificial Life}.
\newblock Addison-Wesley.

\bibitem[Li et~al., 2018]{Li17landscape}
Li, H. et~al. (2018).
\newblock Visualizing the loss landscape of neural nets.
\newblock {\em Proceedings of NeurIPS}, pages 6389--6399.

\bibitem[Liu and Rando, 2011]{Liu11}
Liu, L. and Rando, T. (2011).
\newblock Manifestations and mechanisms of stem cell aging.
\newblock {\em The journal of cell biology}, 193(2):257--266.

\bibitem[Malan, 2021]{Malan21}
Malan, K.~M. (2021).
\newblock A survey of advances in landscape analysis for optimisation.
\newblock {\em Algorithms}, 14(2):40.

\bibitem[Ruddon, 2009]{Ruddon09}
Ruddon, R. (2009).
\newblock {\em Cancer Biology}.
\newblock Oxford University Press.

\bibitem[Southworth et~al., 2009]{Southworth09}
Southworth, L. et~al. (2009).
\newblock Aging mice show a decreasing correlation of gene expression within
  genetic modules.
\newblock {\em PLoS Genetics}, 5(12):pe1000776.

\bibitem[Vijg and Kennedy, 2016]{Vijg16essence}
Vijg, J. and Kennedy, B.~K. (2016).
\newblock The essence of aging.
\newblock {\em Eur Rev Med Pharmacol Sci}, 62(4):381--385.

\bibitem[Williams, 1957]{Williams57}
Williams, G.~C. (1957).
\newblock Pleiotropy, natural selection, and the evolution of senescence.
\newblock {\em Evolution}, 11:398--411.

\bibitem[Yang et~al., 2022]{Yang23epinfo}
Yang, J.~H. et~al. (2022).
\newblock Loss of epigenetic information as a cause of mammalian aging.
\newblock {\em Cell}, 186(2):305--326.

\end{thebibliography}
 
\end{document}